\documentclass[prl,twocolumn,superscriptaddress,amsmath,amssymb]{revtex4} 
\usepackage{appendix}
\usepackage{graphicx}
\usepackage{epsfig} 
\usepackage[dvipsnames]{xcolor}
\usepackage[T1]{fontenc}
\usepackage{wrapfig}
\usepackage[english]{babel}
\usepackage{mathrsfs} 
\usepackage{amsmath}
\usepackage{amsfonts}
\usepackage{bbm}
\usepackage[multidot]{grffile}
\usepackage{multirow}
\usepackage{url}
\usepackage[english]{babel}
\usepackage{subfigure}
\usepackage{soul}
\usepackage{latexsym}
\usepackage{graphicx}
\usepackage{rotating}
\usepackage{hyperref}
\usepackage{amsmath,amssymb,amsfonts}
\usepackage{bm}
\usepackage{color}
\newcommand{\be}{\begin{equation}}
\newcommand{\ee}{\end{equation}}

\newcommand{\ninety}{$90^\circ$ }

\newcommand{\emphasize}{\emph}

\newcommand{\ba}{BaFe$_2$As$_2$}
\newcommand{\se}{KFe$_2$Se$_2$}

\begin{document}
\title{Emergence of novel magnetic order stabilised by magnetic impurities in pnictides}
\author{Carla Lupo}
\author{Thomas Julian Roberts}
\author{Cedric Weber}
\affiliation{King's College London, Theory and Simulation of Condensed Matter, The Strand, WC2R 2LS London, UK}
\begin{abstract} 
The Mermin-Wagner theorem prevents the stabilisation of long-range magnetic order in two dimensional layered materials, such as the pnictide superconductors, unless the magnetism is associated with a discrete symmetry breaking.  A typical known example is the discrete row and column collinear  magnetic state, that emerges in doped iron pnictides materials due to \textit{order-by-disorder} mechanism. In these compounds, the magnetic state competes with superconductivity and the mechanism that stabilizes magnetism remains controversial. In this work, we report the phase diagram of a doped magnetically frustrated Heisenberg model, and the emergence of long-range magnetic order that is stabilized by interactions between the magnetic dopant impurities.
\end{abstract}
\maketitle

%%%%%%%%%%%%%%%%%%%%%%%%%%%
%%%%%%%%%%%%%%%%%%%%%%%%%%%

Unconventional superconductivity occurs in the proximity of magnetically ordered states in many materials \cite{spin_disorder_ref1,spin_disorder_ref2}.
Understanding the magnetic phase of the parent compound is an important step towards understanding the mechanism of superconductivity.
\textcolor{black}{While for cuprates magnetism and the underlying electronic state is understood there is still debate in the case of iron pnictides \ba \cite{spin_disorder_ref3}.} 
Many low-energy probes such as resistivity \cite{spin_disorder_ref_resistivity}, scanning tunnelling microscopy \cite{spin_disorder_ref_nemacity} and angle-resolved photoemission spectroscopy
\cite{spin_disorder_ref_dirac_cone} have measured strong in-plane anisotropy of
the electronic states, but there is no consensus on its physical origin.
It was suggested from first principle calculations \cite{pnictides_orbital_order_and_no_magnetic} that the origin stems from orbital order,
but the obtained anisotropy in the resistivity is opposite to the one found
experimentally \cite{pnictides_wrong_anisotropy_orbital_order}.
A more likely scenario supported by recent neutron diffraction measurements \cite{spin_disorder_incommensurate_magnetism} is related to a spin density wave instability
due to the presence of electron and hole pockets around $\bold{k}=(\pi,0)$ and $\bold{k}=(0,\pi)$. The resulting magnetic order is of nematic type and can be seen as a helicoidal
magnetic state with pitch vector $\bold{Q}=(0,\pi)$ or $\bold{Q}=(\pi,0)$. Recent observation of the collinear magnetic phase has been reported in Mn doped La1111 iron based superconductors \cite{Carretta_2017} induced by the Mn impurities.
The magnetic state induced by Mn and Fe substitutions in F-doped LaFe$_{1-x}$Mn$_x$AsO superconductors, reveals a fast drop of superconductivity and the recovery of a magnetic ground-state at low doping, which have been attributed to RKKY interactions \cite{Gastiasoro_2016}. 
Furthermore in recent studies of optimally electron doped CaKFe$_4$As$_4$ \cite{Fernandes_2018} a novel magnetic order state, called spin-vortex crystal (SVC)\cite{Fernandes_2015,Fernandes_2016}, different from the stripe antiferromagnetic or nematic phase has been observed as the result of the magnetic fluctuations near the $(\pi,\pi)$ \textbf{Q}-vectors.\\
In this work we clarify the interaction of frustrated magnetic systems with impurities and in particular the double-\textbf{Q} state of the canonical $J_1-J_2$ model. 
To describe the low-energy magnetic properties of this system, it has been suggested
early on that a local moment picture may become relevant in the presence of moderately
large electronic correlations\cite{spin_disorder_j2j1}, leading to the Heisenberg model
with both nearest- ($J_1$) and next-nearest ($J_2$) exchange couplings defined by
\begin{equation}\label{ham}
\hat{\cal{H}} = J_1 \sum_{\langle i,j \rangle} 
\hat{{\bf {S}}}_{i} \cdot \hat{{\bf {S}}}_{}
+  J_2\sum_{\langle \langle i,j \rangle \rangle} \hat{{\bf {S}}}_{i} \cdot \hat{{\bf {S}}}_{j},
\end{equation}
in the collinear regime, both $J_1$ and $J_2$ are positive, and $2 J_2 >  J_1$\cite{additional_ref_j1j2_doucot_chandra}.
In this  expression, $\hat{{\bf {S}}}_{i}$ are O(3) spins on a periodic square lattice
with $N=L \times L$ sites. $\langle i,j \rangle$ and
$\langle \langle i,j \rangle \rangle$ indicate the sum over nearest and
next-nearest neighbors, respectively \footnote{$J_1$ sets the energy scale, and
in our work we use $J_2/J_1=0.55$ is used when not specified otherwise, and both $J_1>0$ and $J_2>0$.}.

The first attempt at fitting the experimental spin density wave excitation spectra
with a Heisenberg model suggested that one should use very anisotropic values of $J_1$ ~\cite{pnictides_spin_wave_fit_too_small_J}.
However, it was later shown that the fits of the experimental data included
energy scales beyond 100 meV, which are not well described by magnon excitations~\cite{pnictides_argument_fit_of_nature_is_large_energy}.
A more careful study, including the itinerant character of the electrons\cite{pnictides_ilya_frustration_is_large},
led to the conclusion that pnictides are indeed in the collinear regime with (${\bf Q}=(0,\pi),(\pi,0)$) magnetic instabilities, 
a conclusion supported by first-principle calculations for selenium based compounds (\se) \cite{spin_disorder_electronic_structure}.
We also note that it was also recently argued\cite{heisenberg_biquadratic_pnictide} that 
to get a proper description of magnetic interactions and spin fluctuations in ferropnictides, additional  
biquadratic interactions might be important.

In parallel, it has been suggested both experimentally \cite{spin_disorder_nmr_impurities,bonfa_carretta} and theoretically \cite{spin_disorder_ref_theory_impurities} that 
impurities have a dramatic impact on the magnetic and superconducting properties.
In particular, recent magnetic polarized x-ray measurements suggest that a new type of magnetic
order emerges due to the presence of magnetic impurities in \ba \cite{pnictides_mark_dean_impurities}.
Furthermore, periodic ordering of super-cell structures of vacancies in TlFe$_{1.6}$Se$_2$, observed by electron microscopy, were shown
to induce a spin reorientation in these structures \cite{additional_ref_j1j2_spin_reordering_exp,additional_reference_j1j2_disorder_modulated_lattice}.
All these results, together with results obtained a few years ago on a layered vanadium oxide~\cite{additional_ref_j1j2impurities_careta}, call for an in-depth investigation of the effect
of impurities in this frustrated Heisenberg model. 
%%%%%%%%%%%%%%%%%%%%%%%%%
%%%%%%%%%%%%%%%%%%%%%%%%%

 In this letter we build upon our earlier calculations in Ref\cite{our_paper_pnictites} by extending the calculations to samples doped with both magnetic and non-magnetic impurities, exploring highly doped lattices (up to full doping). In particular, we focus on the competing
magnetic order at high doping, which corresponds to \textcolor{black}{optimally} and \textcolor{black}{overdoped} pnictide samples.
We address the question of the interplay between the frustration induced by the exchange couplings and the disorder induced by the imperfections of the crystallographic
structure. Increasing the doping we expect the possibility of first order phase transitions driven
by a percolation mechanism, where impurities drive local fluctuating order parameters on short distances
and become long range at high dilutions.
%%%%%%%%%%%%%%%%%%%%%%%%%
% Collinear Order Parameter  
%%%%%%%%%%%%%%%%%%%%%%%%%
\begin{figure}[t]
\begin{center}
\includegraphics[width=0.5\textwidth,height=0.25\textwidth]{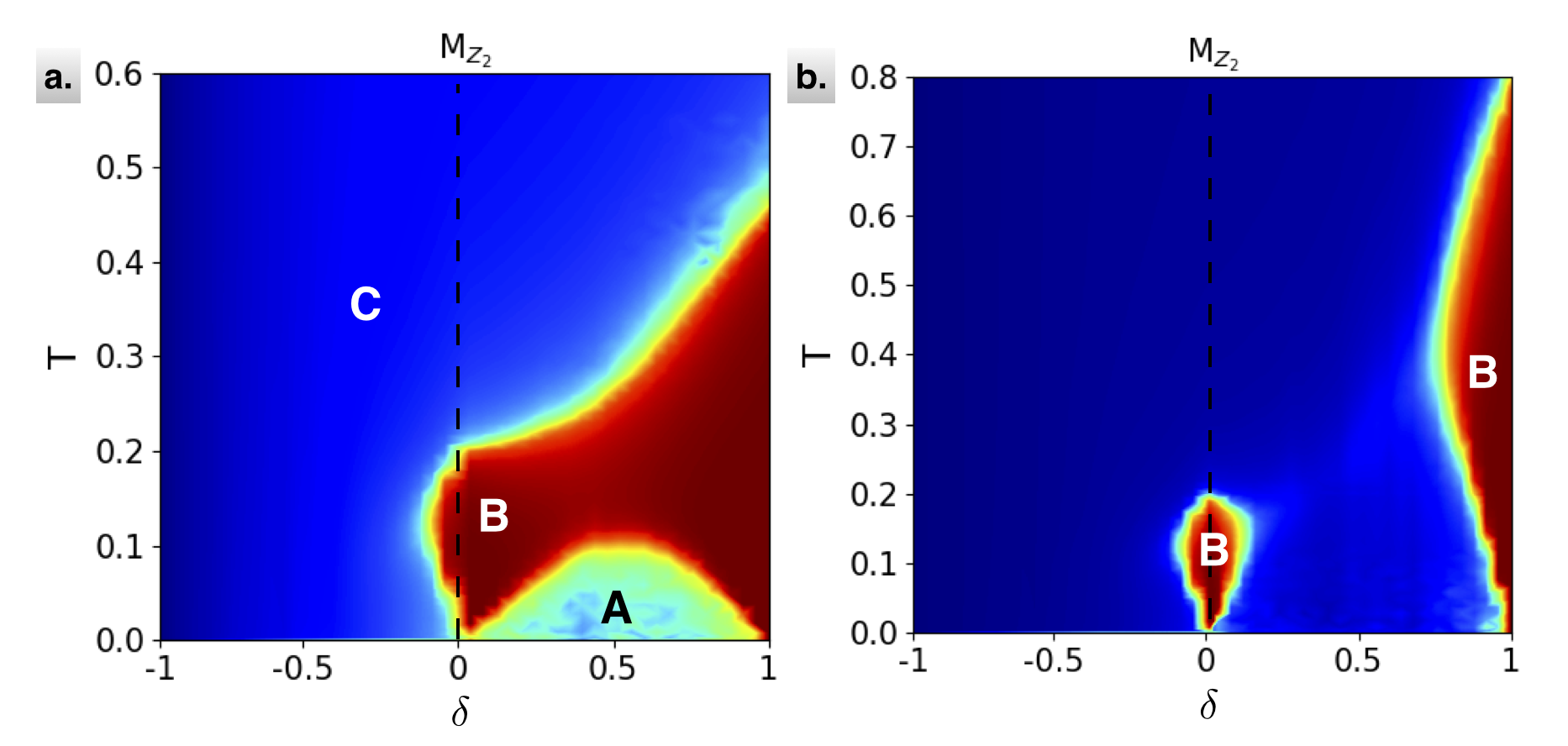}
\caption{(color online) Color maps of the Ising (M$_2$) order in function of temperature and dilution for a $L\times L=50\times 50 $ lattice. Negative and positive dilution refers respectively to doping with non magnetic impurities and with magnetic ones $r=1.5$ (a) and $r=2$ (b). Colours range from blue (minimum) to red(maximum). From low to high temperature different ordered region can be distinguished: A) anticollinear, B) collinear and C) weak N\'eel state.} 
\end{center}
\end{figure}
%%%%%%%%%%%%%%%%%%%%%%%%%
Since density functional calculations, and quite generally quantum based calculations,
are limited to relatively small unit-cells and cannot tackle the issue of large super-cell structures \textcolor{black}{we} limit our calculations to a frustrated classical model \cite{additional_ref_j1j2_frustrated_xy}, and carry out Monte Carlo calculations of the
Heisenberg $J_1{-}J_2$ model in the presence of impurities using the same numerical approach as in Refs.~\onlinecite{cedric,cedric_phonon}.
We limit ourselves to $50\times50$ lattice sizes (2500 correlated atoms), but average over large numbers
of disordered configurations (up to 5000 configurations) by using a BlueGene/Q supercomputer facility.
In \textcolor{black}{the} absence of disorder and at zero temperature, the magnetic vector is ${\bf Q}=(\pi,\pi)$
for $J_2/J_1<0.5$, and  for $J_2/J_1>0.5$
the ground state is continuously degenerate and is characterised
by a bi-partite lattice, with two distinct anti-ferromagnetically ordered
states on \textcolor{black}{each} sub-lattice, with $\theta$ the angle between the two magnetic directions.
\textcolor{black}{
At finite temperature the entropy selection reduces the O(3) symmetry of the ground state to Z$_2$ selecting the states with antiferromagnetic spin correlations in
one spatial direction and ferromagnetic correlations in the other (${\bf Q}=(0,\pi),(\pi,0)$).}
This is the so-called \emphasize{order by disorder} entropic selection \textcolor{black}{and} the associated discrete symmetry
breaking drives a finite temperature Ising-like phase transition \cite{chandra_collin,cedric}. \textcolor{black}{We address how the presence of disorder affects this transition.}

%%%%%%%%%%%%%%%%%%%%%%%%%
% Anticollinear OP
%%%%%%%%%%%%%%%%%%%%%%%%%
\begin{figure}[b]
\includegraphics[width=0.5\textwidth,height=0.25\textwidth]{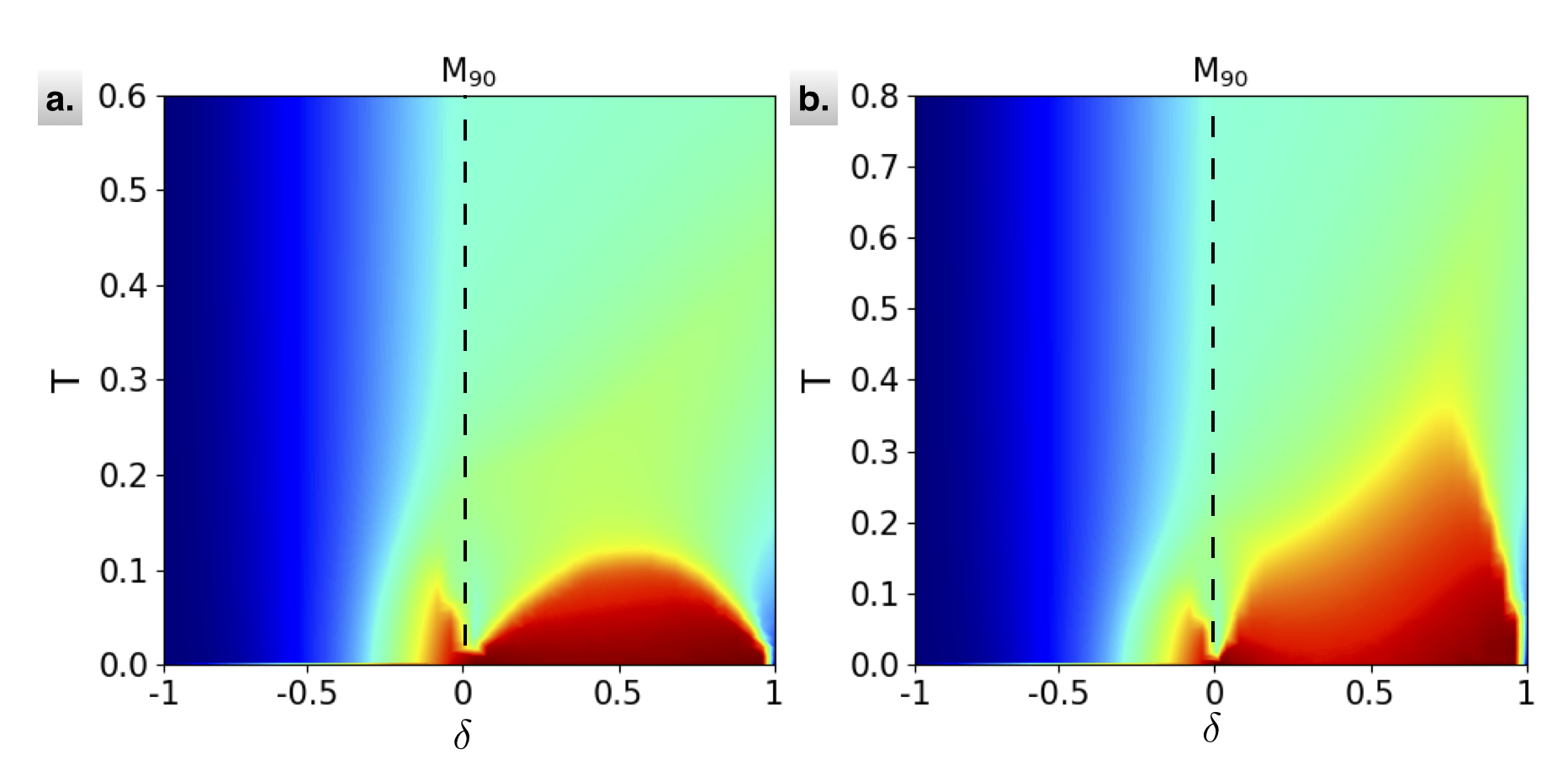}
\caption{(color online) Color maps of the 90 degree (M$_{90}$) order in function of temperature and dilution for a $L\times L=50\times 50 $ lattice. Negative and positive dilution refers respectively to doping with non magnetic impurities and with magnetic ones $r=1.5$ (left panel) and $r=2$ (right panel). Colours range from blue (minimum) to red(maximum).}
\end{figure}
%%%%%%%%%%%%%%%%%%%%%%%%%

We first \textcolor{black}{examine} the phase diagram \textcolor{black}{of both doping regimes where the magnetic moment of the doping (r)} is characterized by its ratio \textcolor{black}{to} the magnetic moment of the undoped compound ($M_{Fe}$ for iron), $r=M_{imp}/M_{Fe}$. The dilution is denoted as $\delta>0$, $\delta<0$  for $r\neq 0$ \textcolor{black}{and $r=0$ respectively}. In Fig. 1 we consider the collinear order parameter constructed from the original spin variables $\hat{{\bf {S}}}_{i}$ 
\begin{equation}
\label{eq:isingvar}
M_2(x)=
(\hat{{\bf {S}}}_{i} - \hat{{\bf {S}}}_{k}) \cdot
(\hat{{\bf {S}}}_{j} - \hat{{\bf {S}}}_{l}),
\end{equation}
where $(i,j,k,l)$ are the corners with diagonal $(i,k)$ and $(j,l)$ of the
plaquette centered at the site $x$ of the dual lattice $[$see Fig. S1(a)$]$, and we define
its normalized counterpart as $Z_2(x)=M_2(x)/|M_2(x)|$.
In this way, the two collinear states with ${\bf Q}=(\pi,0)$ and
${\bf Q}=(0,\pi)$ can be distinguished by the value of the
Ising variable, $Z_2(x)= \pm 1$.
For impurities with a 50\% larger magnetic moment (see Fig. 1(a), $\delta>0$), we observe that there exists a temperature range $T=(0.1,0.2)J_1$ where the \textcolor{black}{collinear} order survives at all \textcolor{black}{dilutions}. \textcolor{black}{However,} the transition from \textcolor{black}{collinear} to paramagnetic (from region B to C) \textcolor{black}{at high} temperature \textcolor{black}{increases} from 0.2 to 0.45. This can be explained by a very simple argument; in the fully doped regime \textcolor{black}{all} spins are $1.5$ \textcolor{black}{times} larger and so the energy scales are rescaled by a factor $1.5^2$, increasing T$_c$ in turn by a factor $2.25$. Differently from the case with $r=1.5$, we observe that the Ising\textcolor{black}{-like} order is rapidly suppressed by doping with non-magnetic $r=0$ impurities (Fig. 1(a), $\delta<0$) or impurities with a large magnetic moment $r=2$ (Fig. 1(b), $\delta>0$)\textcolor{black}{, with no collinear} magnetic order obtained beyond $8\%$ \textcolor{black}{dilution}. This is expected for the case of non-magnetic dopants, where large dilutions prevents the propagation of long-range magnetic order \textcolor{black}{as} the magnetic order propagates by short-range correlations. The quenching of low energy fluctuations upon the introduction of non-magnetic impurities have been observed \textcolor{black}{experimentally} both in vanadates \cite{carretta} and pnictides \cite{bonfa_carretta}.
Although the Ising order survives for $\delta>0$, if we look at \textcolor{black}{ fixed dilution, $\delta=50\%$,} we observe that the \textcolor{black}{collinear} order is also suppressed at low temperature (region A), and we obtain a re-entrance transition of the \textcolor{black}{collinear} order (region A to B). This is expected at low temperature and low doping\textcolor{black}{; }it has been shown that around a single impurity the degeneracy of the ground-state of the $J_1-J_2$  model is lifted and the 
\ninety magnetic order is selected from the manifold by an energy optimization process \cite{henley_j_applied_87_paper_def_anticollinear,henley_j1j2_dilution}. Note that this latter mechanism is driven by an energy optimization and is not expected to survive \textcolor{black}{to} high temperatures.
In Fig. 2a,b we report the anticollinear order 
\begin{equation}
    M_{90}(x)=\vert (\bf{\hat{S}}_i-\bf{\hat{S}}_k)\times (\hat{\bf{S}}_j-\hat{\bf{S}}_l )\vert 
\end{equation}
where $(i,j,k,l)$ defines the same plaquette as in Eq.\ref{eq:isingvar}  $[$see Fig. S1(a)$]$.

Our results confirm that the order stabilized in region $A$ in Fig. 1a is the \ninety order. Local fluctuations of the \ninety order around impurities percolate and form a stable order at  low temperature. At \textcolor{black}{high} temperature the entropic contribution dominates  and the Ising-like order is recovered (Fig 1.a, $\delta>0$). 
Note, however, that if the magnetic moment of the dopant is large ($r=2$), the entropic contributions aren't able to recover the \textcolor{black}{collinear} order and the \ninety order surprisingly stabilizes at high temperature until the paramagnetic phase is obtained (Fig. 2b, $\delta>0$ and Fig. 1.b $\delta>0$), leading to a suppression of the Ising order in between the undoped and fully doped regions. This has been observed in the superconducting pnictides doped with Ir \cite{pnictides_mark_dean_impurities} where the collinear order is suppressed when the  dilution is greater  than $\delta>0.047$. This is in agreement with the quenching of the collinear phase  observed as impurity ratio $r=2$ in the doping region $\delta=[0.2, 0.8]$.   Indeed in Fig. 2b we can clearly see that at approximatively half doping  the low temperature range is fully dominated by the anticollinear order $M_{90}$ being the \textcolor{black}{collinear} order, $M_{Z_2}$ equal to zero Fig. 1b. 
Note that this mechanism is not obtained by doping with non-magnetic impurities (Fig. 1a $\delta<0$ and Fig. 2a $\delta<0$), as the suppression of the Ising\textcolor{black}{-like} order is not concomitant with the stabilization of a competing order.
In the dilution range $\delta<0.2$ and $\delta>0.8$, the competition between the entropic and the energetic contribution is restored and interestingly we observed that the re-entrance transition (region A to B, Fig 1b, 2a) is characterized by a sharp cross over. This mechanism is rationalized in Fig. 3a, where we considered a single impurity case. We observe that the cross-over is characterized by a magnetic phase (Fig 3c) different from both the anticollinear (Fig 3b) and the collinear case (Fig 3d). This intermediate phase consists of two distinct antiferromagnetically ordered  states on two sublattices with a relative angle $\alpha$ between their magnetization axis which is selected by the impurity spin direction. 
%%%%%%%%%%%%%%%%%%%%%%%%%
% SINGLE IMPURITY ALL
%%%%%%%%%%%%%%%%%%%%%%%%%
\begin{figure}[t]
\begin{center}
\includegraphics[width=0.5\textwidth,height=0.55\textwidth]{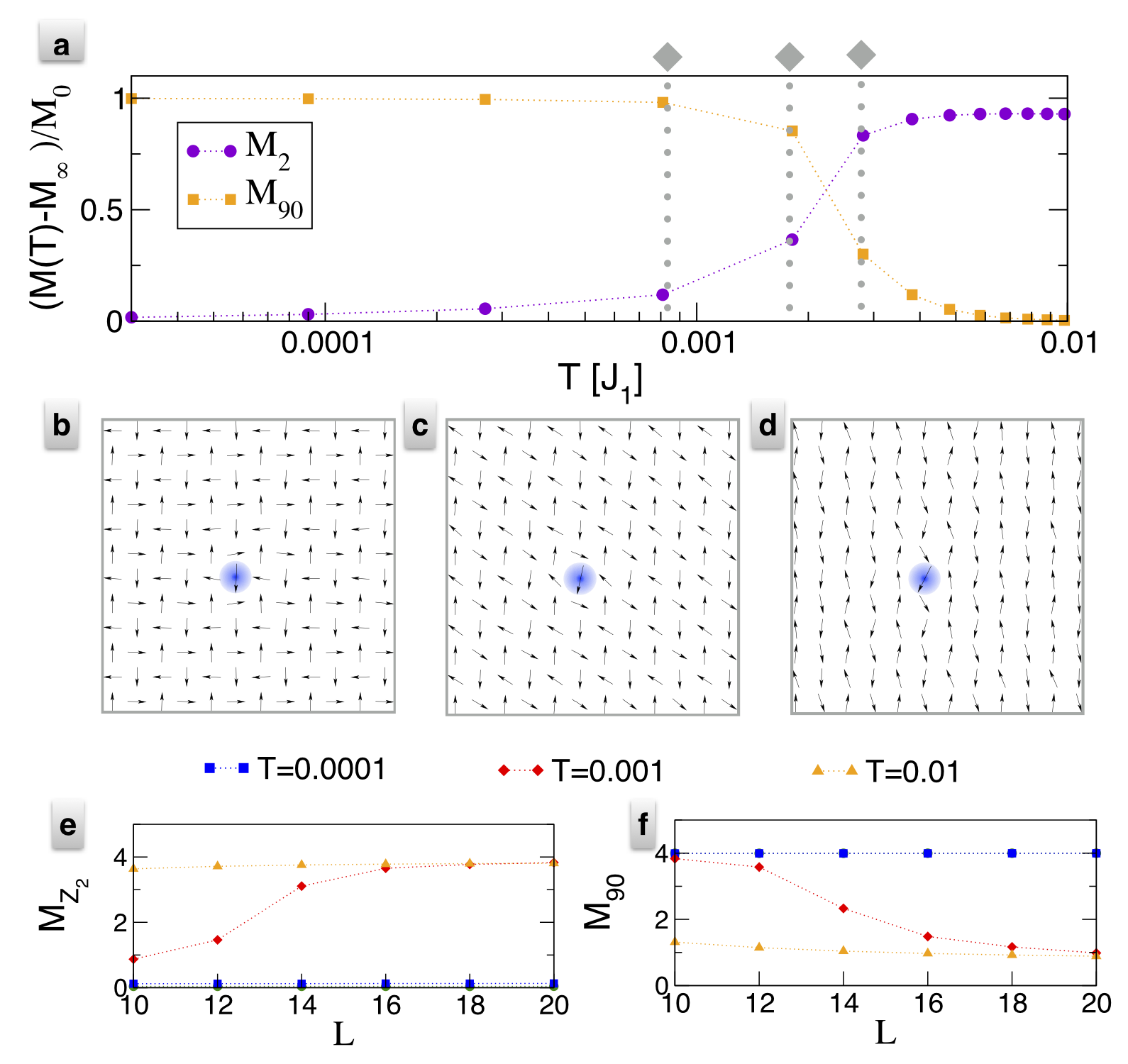}
\caption{Single magnetic impurity case with $r=1.5$. a) Temperature dependence of $M_2$ and $M_{90}$ for a lattice with $L=12$.  b)-c)-d) Typical spin configuration obtained at three fixed temperatures at $T=0.0008, 0.00181, 0.00281$ $ J_1$. e)- f) respectively  collinear and anticollinear order parameter values for fixed temperatures $T=0.0001,0.001,0.01 $\textcolor{black}{$J_1$} in function of different lattices with linear dimension $L$ .  }
\end{center}
\end{figure}
%%%%%%%%%%%%%%%%%%%%%%%%%
%%%%%%%%%%%%%%%%%%%%%%%%%
This suggest that there is a crossing of the free energies of the \ninety and \textcolor{black}{collinear} orders at the transition, where the competition in the free energies $F=E-T*S$ happens between the energy term $E$ and entropic contribution $T*S$. As this process is very much dependent on the local disorder configurations, the temperature associated with the sharp cross-over is also dependent on the disorder configurations. In an experiment, or in our computed physical observables which are averaged over large disorder samples, the transition is a smooth cross-over, hiding the physical explanation related to the competition of energetic and entropic terms. 
Remarkably, the mechanism which determines the \textit{energy vs entropy} competition  is different respective to vacancies or magnetic doping. Indeed in case of non magnetic impurity the transition between the anticollinear and the collinear phase is happening through a coexistent phase:  it was shown (Ref \cite{our_paper_pnictites}) at finite temperature the anticollinear order stabilizes locally around the impurity and with the collinear states recovered outside this region. Instead, in case of magnetic impurities, we observe that the magnetic phase which characterizes the crossover is not a coexistent phase of collinear and  anticollinear order. 
The transition between the \ninety and \textcolor{black}{collinear} order is rationalized with respect to the lattice size in Figs 3a-b, where we show both the order parameters at three different temperatures, $T=10^{-4},10^{-3},10^{-2} J_1$, for a single impurity embedded in a lattice of size $L=(10, 20)$. Note that periodic conditions are used in this simple model, such that the lattice size mimics the average distance between impurities at high dilutions.  At low temperature $T=10^{-4}$, as entropic contributions are absent, we observe that the \ninety order dominates as expected for all cases (analytic argument at  $T=0$ in suppl mat. Fig S1b). As temperature is increased to $T=10^{-3}$ and $T=10^{-2}$, we observe that the \ninety degree is stabilised at small L, but the \textcolor{black}{collinear} order wins in larger lattices where the entropic contributions in turn become larger. This illustrates the mechanism obtained around the large dilution (small L), where the \ninety order is stabilized, and at low dilutions (large L), where the \textcolor{black}{collinear} order wins. 

Further insights about the transition between the different magnetic phases is shown in Fig. 4. For doping with magnetic impurities (Fig. 4a-b), we obtain as expected a large peak in the specific heat at the transition associated with the loss of the \textcolor{black}{collinear} order (region B to C, Fig. 1a $\delta>0$). As we do not observe a drop in the specific heat along the Ising\textcolor{black}{-like} transition in Fig 4a (where $r=1.5$), we conclude that the transition remains second order along this line. Surprisingly a continuous transition occurs also in Fig 4b (with $r=2$) at the transition between the anticollinear and the paramagnetic phase, even if the Ising-like order is zero for all $T$. 
In more detail, at fixed dilution $\delta=25\%$ we observe that the  melting of the anticollinear order occurs with a  cross over associated with a non divergent peak of the specific heat (Fig. 4d).  Note that the specific heat also indicates fluctuations at the re-entrance transition
(region A to B). Thus for a critical $r_c<2$, the intersection of the different magnetic phases turns into three-critical crossing points. 
 
For doping with non-magnetic impurities we observe the irising of the peaks for the Ising-like transition (fixed low doping) which is consistent with what observed so far in case of magnetic impurities. A more interesting and novel behaviour is observed at fixed low temperature where there exists a continuous pathway which does not involve any sharp transition. This is crucial for applications because it does not involve any energy cost. This was not observed in the previous work because no fluctuations where considered. At zero temperature 
We observe that there are no energy fluctuations associated with the percolation transition which is instead indicated by the sudden disappearance of the susceptibility
at $8\%$ in Fig. S5 (suppl mat) typical of a first-order transition.

%%%%%%%%%%%%%%%%%%%%%%%%%
% Fig: SPECIFIC HEAT
%%%%%%%%%%%%%%%%%%%%%%%%%
\begin{figure}
\begin{center}
\includegraphics[width=0.5\textwidth,height=0.45\textwidth]{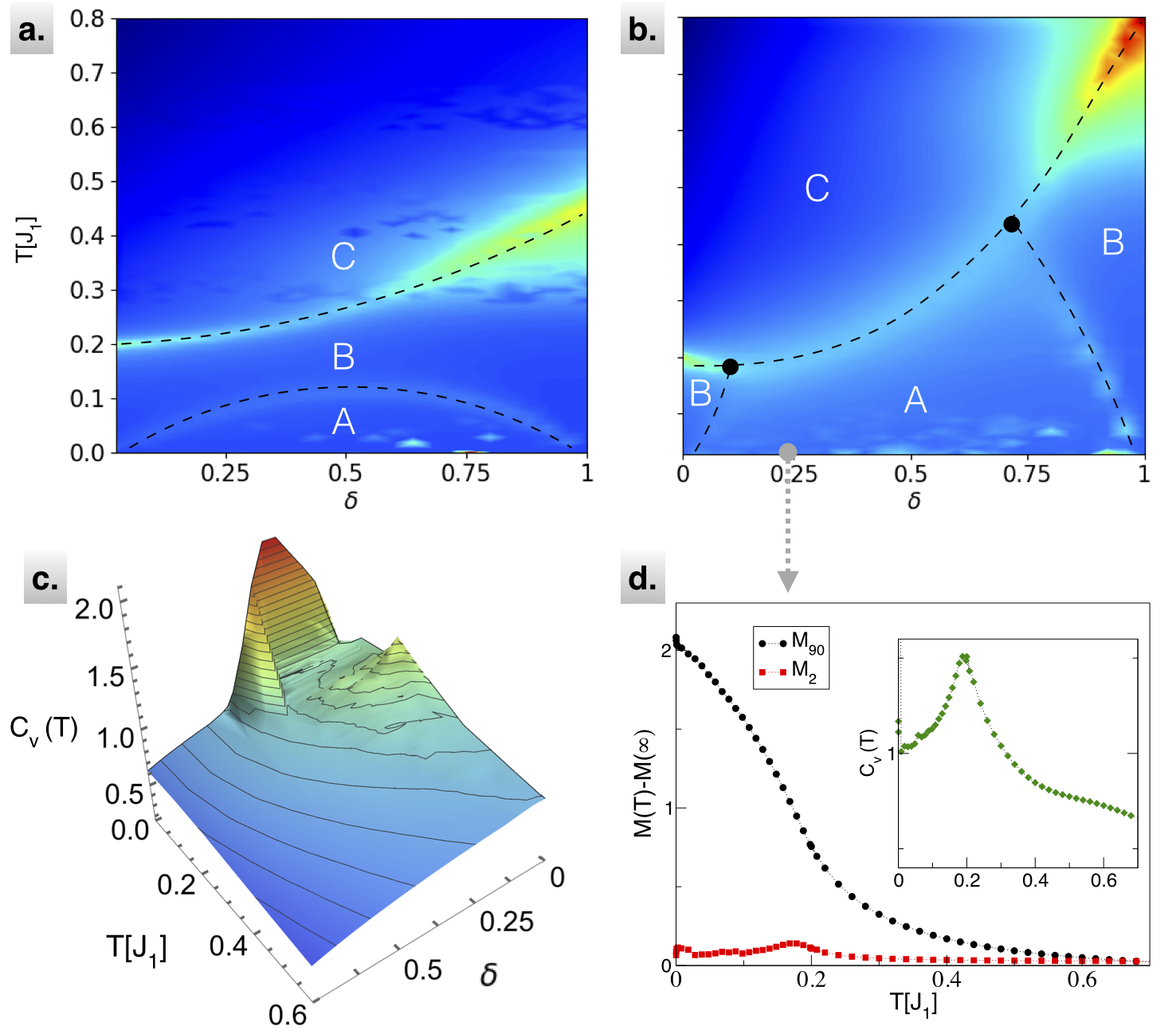}
\caption{(color online) Color maps of the specific heat in function of temperature and dilution for a $L\times L=50\times 50 $ lattice respectively for  magnetic impurities $r=1.5$ (a) and $r=2$ (b) . Colours range from blue (minimum) to red(maximum). Black dashed lines are guide to the eyes to distinguish the three ordered states: A (anticollinear), B (collinear) and C (paramagnetic). }
\end{center}
\end{figure}
%%%%%%%%%%%%%%%%%%%%%%%%%
%%%%%%%%%%%%%%%%%%%%%%%%%

%--------------------------------------
% Par: CONCLUSION
%--------------------------------------
In conclusion we found that the order by disorder entropy selection, associated with the Ising-like phase transition that appears for $J_2/J_1>1/2$ in the pure spin model, is quenched at low temperature due to the presence of impurities. Indeed, irrespective of the magnetic ratio of the dopant an anticollinear order is stabilized around the impurities, which in turn induces a reentrance of the Ising-like phase transition. The melting of the collinear order occurs via two different mechanisms: i) through a percolation transition from increasing dilution (at fixed temperature) and ii) via a sharp cross-over due to the energetic versus entropic contribution increasing temperature (at fixed doping). While the former exists irrespective to the nature of the dopant the latter is highly affected by the ratio of the magnetic impurities. Remarkably we identify a regime where the anticollinear order is stabilized at finite temperature without going through the collinear phase.

%###########################################################
%###########################################################
%###########################################################
%###########################################################
\paragraph{Acknowledgments } We thank R. Fernandes and P. Carretta for quite insighful suggestions following their critical reading of the manuscript.
C.L. is supported by the EPSRC Centre for Doctoral Training in Cross-Disciplinary Approaches to Non-Equilibrium Systems (CANES, EP/L015854/1). C.W. gratefully acknowledges the support of NVIDIA Corporation, ARCHER UK National Supercomputing Service. We are grateful to the UK Materials and Molecular Modelling Hub and the Hartee Centre (Bluegene -Q) for computational resources, which is partially funded by EPSRC (EP/P020194/1).

%\bibliographystyle{apsrev}
%\bibliography{mybiblio_phd_source.bib}

\newpage
\newpage
\setcounter{figure}{0}
\setcounter{equation}{0}
\renewcommand{\thefigure}{S\arabic{figure}}
\renewcommand{\citenumfont}[1]{#1}

\appendix

\begin{center}
\textbf{Supplementary Material for\\ ``Magnetic impurities in the frustrated Heisenberg model"}
\end{center}

\section{Dependence on the spin impurity ratio}

We would like to investigate the energy gain obtained by the distortion of an angle $\alpha$ around a single impurity with magnetic magnitude $r$. Since we assume that the anticollinear order is stabilized beyond the next nearest neighbours of the impurity - only the first nearest neighbours are tilted as shown in Fig. S1(a)- we have that the local energy is:
    \begin{equation}\label{eq:energy_ki_1}
\begin{split}
   E(\lambda=J_2/J_1, \alpha, r)/J_1&=4\cos(\pi/2-\alpha)+8\lambda\cos(\pi-\alpha)\\
   &+4\lambda\cos(\pi-2\alpha)+8\cos(\pi/2+\alpha)\\
   &\textcolor{black}{-}12\lambda+4r\left(\textcolor{black}{-}\lambda+\cos(\pi/2-\alpha)\right) 
\end{split}
\end{equation}
As discussed in Ref.\cite{our_paper_pnictites} the energy gain at zero temperature decreases monotonically when $J_2/J_1$ increases. The value  of $\lambda=J_2/J_1$ above which this energy contribution becomes small depends on the magnitude of the magnetic impurity. 
Starting from the undoped regime $r=1$ we expect the energy contribution to be zero.  This is explained by the fact that the ground state is described by two sublattices, continuously degenerate with respect to one another, and which have a relative angle $\theta$ between their magnetization axis this is the case at $T=0$.

%%%%%%%%%%%%%%%%%%%%%%%%%
% Analytical argument for T=0 and single impurity
%%%%%%%%%%%%%%%%%%%%%%%%% 
\begin{figure}[h]
\begin{center}
\includegraphics[width=0.47\textwidth,height=0.2\textwidth]{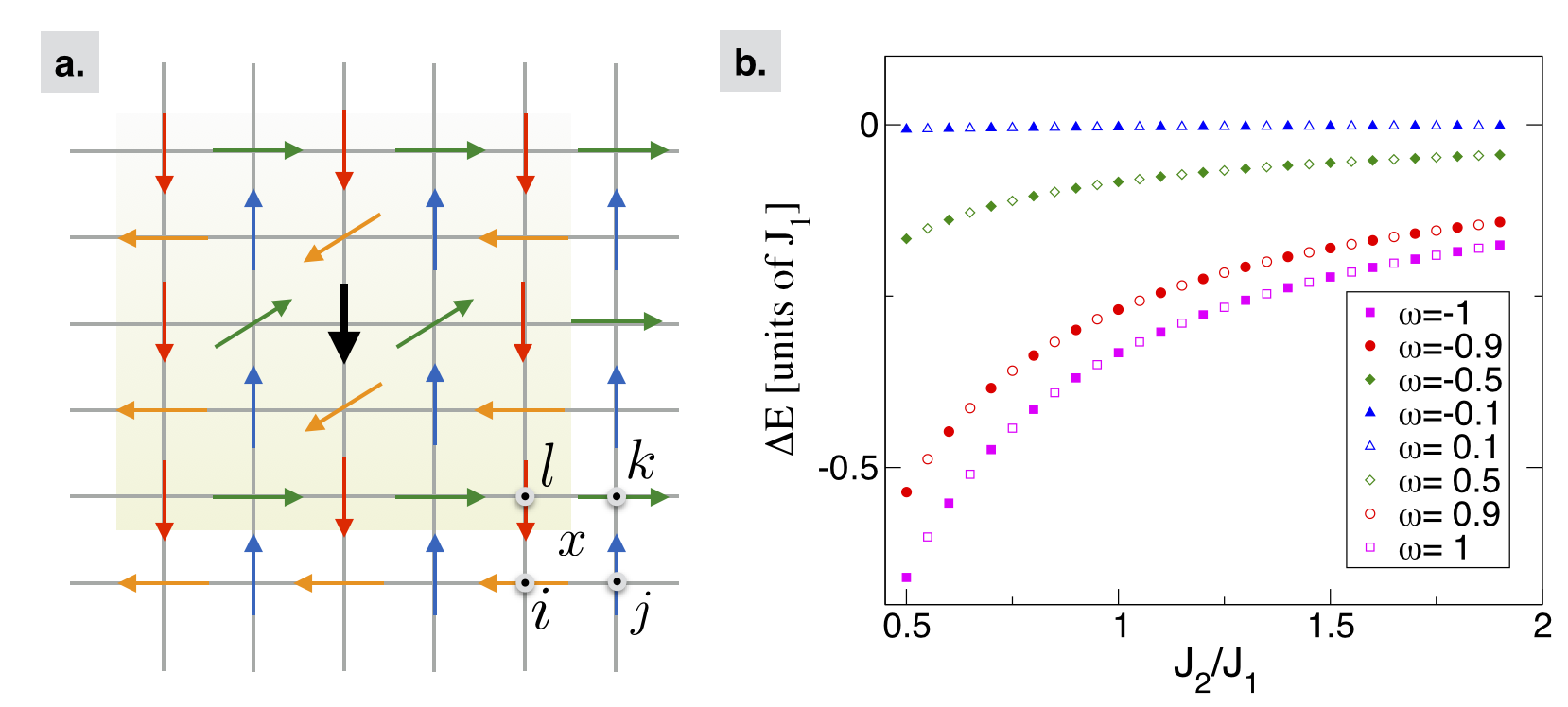}\label{fig:ki_1_aa}
\caption{Single impurity case. a) local distortion of the nearest neighbours around the impurity (black arrow) of an angle theta and \ninety order outwards. Plaquette representation centered in $x$ with corners $(i,j,k,l)$ on the left bottom side of the picture. b) Energy gain $E(\alpha)-E(\alpha=0)$ at zero temperature as a function of $J_1/J_2$ computed with the variational argument in Eq.\ref{eq:energy_ki_1} where only the nearest-neighbour sites of the magnetic impurity are distorted by an angle $\alpha$. Different symbols relate to different spin impurity ratio $r=1-\omega$.} 
\end{center}
\end{figure}
%%%%%%%%%%%%%%%%%%%%%%%%%
We observe that the energetic optimization is symmetric  with respect to $\omega=1-r$.

\section{Simulation Methodology: heatbath and parallel tempering}
\begin{figure}[t]
\begin{center}
\includegraphics[width=0.5\textwidth,height=0.2\textwidth]{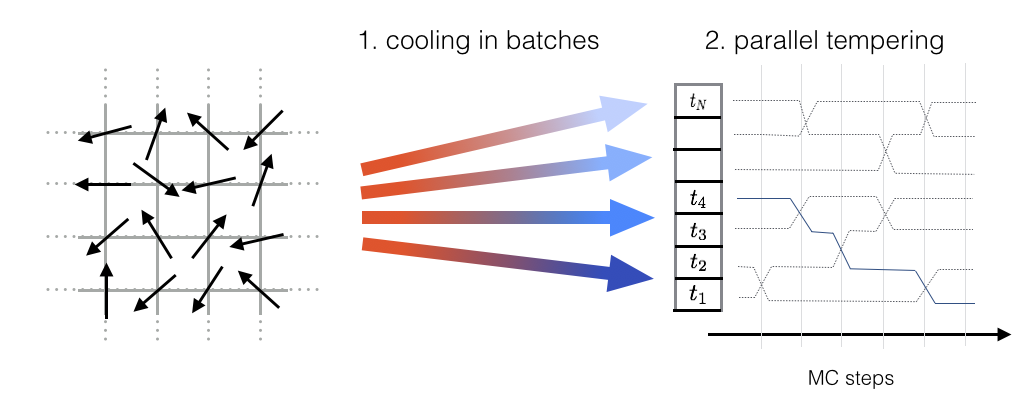}\label{}
\caption{Diagram of the simulation procedure. Starting from a random configuration, we cooled the system in N replicas at adjacent temperatures $\{t_1,\cdots,t_N\}$. We then perform parallel tempering starting from these N configurations:  an exchange of configuration information is allowed between replicas at adjacent temperatures to prevent to system from being trap in a local minimum. There is a further parallelization on top of this process which starts from different iniziatila conditions and It is crucial in the final  averages evaluation. }
\end{center}
\end{figure}

Our Monte Carlo simulations were performed by a two-part process with both parts involving the use of a heat bath algorithm Ref.\cite{Miyatake_1986} to generate new spin configurations. The first part involved a shorter cooling phase from a high temperature. 
This process allowed the initially random configurations to slowly approach equilibrium configurations and prevent the system from being trapped in local minima at low temperature.\\
The second part was the main MC stage where the temperature was kept fixed. In particular to get a better exploration of the phase space we implemented a parallel tempering algorithm.\\
Parallel tempering  (Ref.\cite{Swendsen_1986}) involves simulating a number of replicas simultaneously and allowing configurations to be swapped between adjacent temperatures while performing Monte Carlo steps between swaps. Replicas are allowed to switch based upon the condition of detailed balance. To enforce this we used the standard Metropolis Monte Carlo condition to decide if swaps between replicas at different temperatures were accepted. 
\begin{equation}
    \omega_{m\to n}=\begin{cases} 
    1 \quad\text{if}\quad E_m\leq E_n\\
    exp\left(-(E_n-E_m)/KT\right) \quad\text{if}\quad E_m> E_n
    \end{cases}
\end{equation}
To achieve the best possible sampling at low temperature we consider
batches with temperatures in a geometric progression ($T_i/T_{i-1}=constant$).Since the computational effort increases on the order of the number of the replicas $N$, it became crucial the use of large CPU clusters. A further parallelization was used also on top of the schematic representation in Fig. S2 in order to get better averages of the observables. All our calculations and order parameters were measured during this second stage taking ensemble averages across all spin configurations at a particular temperature. 
The method originally limited to problems in statistical physics, it was later apply to Monte Carlo simulations of biomolecule. The variety of fields in which it has been generalized include polymers \cite{Doxastakis_2004}, protein \cite{Im_2004} and spin glasses \cite{Katzgraber_2001}.

\begin{figure}[t]
\begin{center}
\includegraphics[width=0.5\textwidth,height=0.3\textwidth]{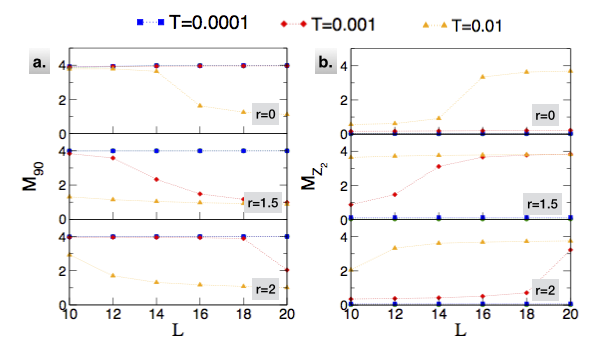}\label{fig:ki_1_sim}
\caption{Single impurity case. Anticollinear $M_{90}$(left) and collinear $M_2$ (right) order obtained by Monte Carlo at fixed $T=0.00001,0.0001,0.001,0.01 $\textcolor{black}{$J_1$}, in case of single magnetic impurity with magnitude $r=0,1.5,2$ in function of different lattices with linear dimension $L$.  }
\end{center}
\end{figure}
%%%%%%%%%%%%%%%%%%%%%%%%%%%%%

%%%%%%%%%%%%%%%%%%%%%%%%%%%%%
% CV doping with vacancies
%%%%%%%%%%%%%%%%%%%%%%%%%%%%%
\begin{figure}[t]
\begin{center}
\includegraphics[width=0.35\textwidth,height=0.3\textwidth]{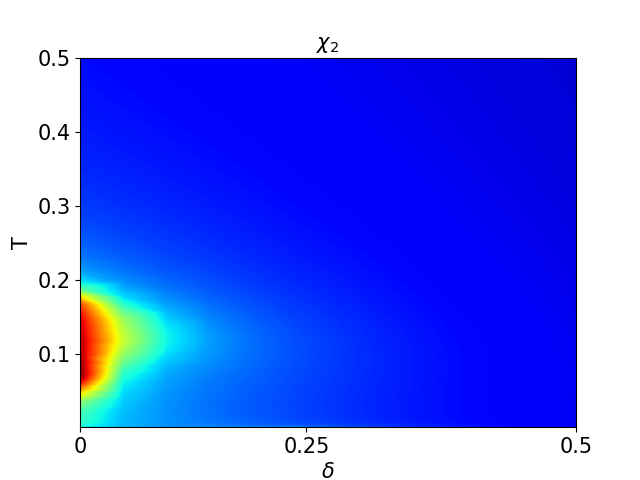}\label{fig:ki_1_sim}
\caption {(color online) Color map of the fluctuations $\chi_{M_2}$} of the collinear order parameter for the case of doping with vacancies ($r=0$). 
\end{center}
\end{figure}
%%%%%%%%%%%%%%%%%%%%%%%%%%%%%

\end{document}